\documentclass[a4paper,fleqn]{mnras}

\usepackage{newtxtext,newtxmath}
\usepackage[T1]{fontenc}
\usepackage{ae,aecompl}
\usepackage{graphicx}   % Including figure files
\usepackage{amsmath}    % Advanced maths commands
\usepackage{amssymb}    % Extra maths symbols

\title[Crab Nebula]{High Energy emission from Crab Nebula}

\author[Tevdorashvili]{
Beka Tevdorashvili$^{1}$
\\
$^{1}$Centre for Theoretical Astrophysics, Institute for Theoretical Physics, Ilia State University, Tbilisi 0162, Georgia\\}

\pubyear{2020}

\begin{document}
\label{firstpage}
\pagerange{\pageref{firstpage}--\pageref{lastpage}} \maketitle

% Abstract of the paper
\begin{abstract}
Flaring episodes from Crab Nebula have been observed. A new mechanism of emission is explored. Particles in Crab pulsar are accelerated to multiple Tev energies, by some mechanisms, described in the paper and they are the reason of observed emission. We argue, that after accelerating to high energies, they may maintain the force-free regime - moving along the magnetic field lines in such a way that no force acts on them. This way, they can move to the Nebula without loss, which makes Tev energy emission from Crab Nebula quite possible.
\end{abstract}


\section{Introduction}

There has been very interesting observations made lately about the high energy emission from the Crab Nebula \cite{aliu2013search}. Inverse-Compton mechanism for low-energy particles and synchrotron radiation mechanism for relativistic particles have been considered as the explaining mechanisms for the observed emission. 
It has already been shown, that particles are able to gain their energies up to 100s of Tev, or even Pev orders \cite{mahajan2013ultra}. As the Crab pulsar is considered as the energy source of the Crab Nebula emission \cite{staelin1968pulsating}, it is very interesting to understand the whole process of the particle motion. 

To begin with, we are going to consider the possible mechanisms, which can explain the particle acceleration in Crab pulsar. This process, as mentioned above, should become the beginning of the process, which will, in the end, result in radiation from Nebula. Next, we will try to describe particle motion through Nebula.

We argue, that in some special cases, these high energy particles can leave the light cylinder and enter the Nebula being on force-free regime. These regime can be maintained by the particles moving on Archimedes spirals \cite{rogava2003centrifugally}. This very special case, is needed to explain, why particles accelerated in Crab pulsar, emit from the Nebula the same amount of energy as they posses in pulsar surroundings.

To understand the phenomenon thoroughly, it is also very important to understand the emission mechanism from the Nebula. Despite the huge amount of theoretical studies in this field, there is no wholly accepted mechanism explaining the high energy emission from Crab Nebula. 

In the first section we will discuss the possible particle acceleration mechanisms in Crab Pulsar. In this section we are going to estimate the energies particles can maintain, considering several models explaining the maintained energies. Next we will estimate the approximate radiated energy and finally we will consider, that particles move on Archimedes spirals and discuss the possible radiation from Crab Nebula, considering aforementioned factors.

\section{Particle acceleration in pulsar}
We consider, particles moving in force-free regime, spend their energy supply in the Nebula. Energy transfer from pulsar to the Nebula has already been studied before \cite{machabeli1996problem}. We argue, that a few amount of particles with Tev energy, enter the vicinity of Nebula without any loss and radiate there. This consideration quite well explains the observed spectrum of the Crab Nebula \cite{de1996gamma}. Crab Nebula radiation spectrum maximum is at gamma range. But, for our supposition, very interesting part of the spectrum begins at $10^{2}$ Mev, when the spectrum curve starts to grow and picks the maximum at Tev energy.

It has been shown, that pulsar rotational energy could be converted into the energy of Langmuir waves \cite{machabeli2015origin}. And, as the electromagnetic waves do not damp on particles, which is the result of phase velocity of perturbations having higher velocity, than the speed of light, they create plasmon condensate, which reaches the Crab Nebula, without any loss. 

The strength of the wave field as a number, was introduced in 1971 \cite{gott1971astrophys} as dimensionless Lorentz invariant parameter $Z = \frac{{e E {\lambda} }}{{m c^{2}}}$. Where, $e$ and $m$ are the charge and the mass of electron, $E$ is the electric field magnitude, $\lambda$ is the wavelength and $c$ is the speed of light. It was suggested, that in the center of Crab pulsar magnetosphere, strength of the wave field is very low ($Z<<1$), but it takes it's much higher quantity in the vicinity of light cylinder ($Z>>1$). In the same year synchro-Compton mechanism of radiation was suggested \cite{rees1971new}. As the electric and magnetic fields become linearly dependent on the distance, thus strength of the wave field $Z - r^{-1}$. Debye radius of the relativistic electron is $R_{D}=\frac{c \gamma}{\omega_{B}}$. Electron travels the distance along the magnetic field line  $R_{E}=\frac{R_{E}}{\gamma}=\frac{c}{\omega_{B}}$. And thus, for the wave length $\lambda=\frac{c}{\omega}>>R_{E}$. Which makes wave magnetic field homogeneous. Dipole emission generates far from the magnetic dipole. So the emission may occur outside of the magnetosphere, in the vicinity of light cylinder.

It quite well explained, gamma and x-ray radiation.
Still, these aforementioned mechanisms could not explain, the reason, why is the maximum of Tev radiation, lower than the maximum of Gev radiation in Crab Nebula radiation spectrum.

In 1975 Cocke \cite{cocke1975coherent}, suggested, that particles, moving along the curved field lines and radiating high energies, could not be accelerated to sufficiently ($\gamma = 4 * 10^{6}$) in Crab pulsar magnetosphere. This consideration came into agreement with a bit earlier work done by Tademaru \cite{tademaru1973energy}. Thus, the logical scenario, for the particles, was to accelerate to such energies in Crab Nebula. 
To sum up their work, it was believed, that high energy radiation of particles, while motion on curved magnetic field trajectories, could be maintained ($100 Kev$), in a case, where particles have Tev range energy. It was suggested, that such high energies could not be maintained in the magnetosphere.
This suggestion makes it possible, that particle in Crab Nebula have energies up to Pev range. Aforementioned synchro-Compton mechanism of radiation could possibly lead to such scenario.

According to \cite{mahajan2013ultra} a very big amount of low Lorentz factor particles in the bulk of e-p plasma, are responsible to produce Langmuir waves.  The distribution function can be seen on Fig 3. It shows, that the amount of high Lorentz factor particles is much less than the amount of low Lorentz factor particles. Thus, Langmuir waves, which are produced by low energy particles, are later Landau damped on electron beam particles, which have very high Lorentz factors. 

Reason of beam particles having high energies and thus, Langmuir waves damping on beam particles is easy to understand, as dispersion of electrostatic waves in relativistic e-p plasma is allowed in a limit, when phase speed of the wave approaches the speed of light $v_{ph} = c$. According to (Lominadze) dispersion relation is
\begin{equation}
\omega = k c - \beta (k - k_{0})
\end{equation}
In a case, where phase velocity and the speed of light are the same, $ \beta $ approaches to zero. Which, on the other hand means, that Lorentz factor needs to be as high as possible. This is the reason, why Langmuir waves are damped on high Lorentz factor beam particles.

It is believed, that energy of a beam electron can get as very high 100s of Tevs or higher \cite{mahajan2013ultra}. Particles moving along the magnetic field lines gain energy provided by reaction force, with radial and azimuthal components given by \cite{rogava2003centrifugally}

\begin{equation}
F_{r} =  - \frac {\Phi} {\sqrt{1 + \Phi^2}} {\bf F}
\end{equation}
\begin{equation}
F_{\phi} =   \frac {\Phi} {\sqrt{1 + \Phi^2}} {\bf F}
\end{equation}

Where, $\bf F$ is reaction force and $\Phi$ is defined as 

\begin{equation}
\Phi = r \phi' (r)
\end{equation}

with $\phi$ being curve, that particle moves along. For the particles energies gained by reaction force
\begin{equation}
E=\frac {n_{1} F_{reac} \delta r} {n_{GJ}}
\end{equation}

After estimations, using parameters near light cylinder $n_{GJ} = 2 * 10^{7} cm^{-3}$ and $n_{1} = 2.5 * 10^{11} cm^{-3}$ particles energy needs to be 100s of Tev.

\begin{figure}
	\includegraphics[width=\columnwidth]{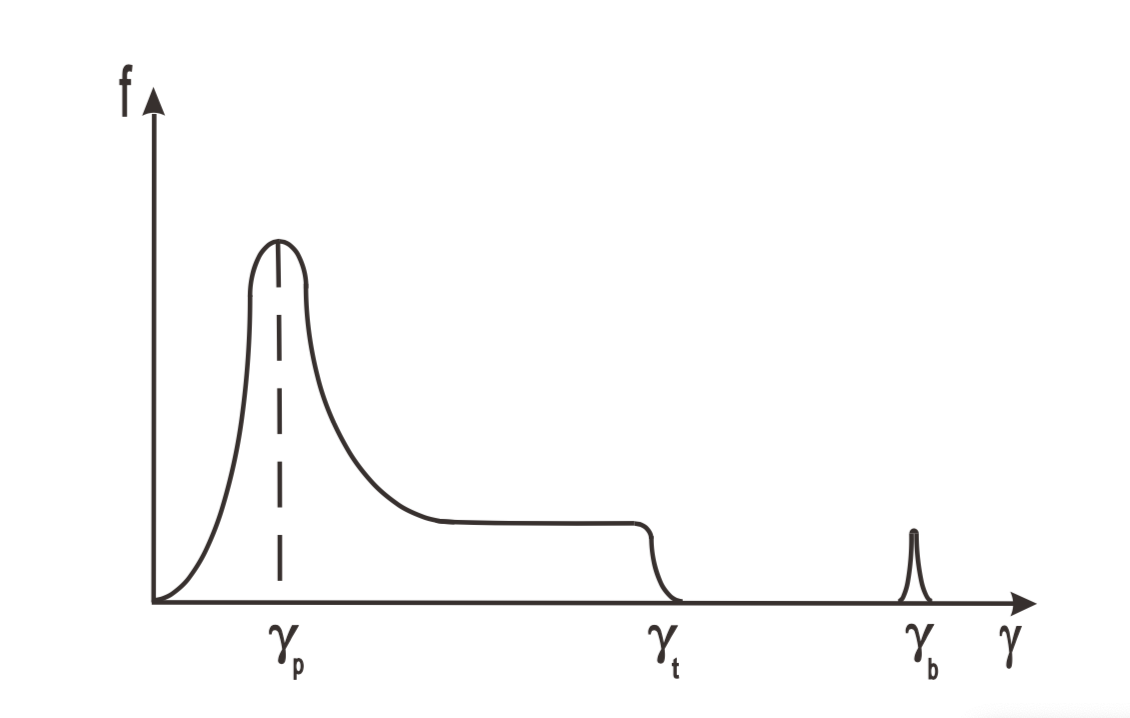}
    \caption{As it is shown on the figure, low Lorentz factor e-p plasma particles are much more, than the bulk particles, with much higher Lorentz factors}
    \label{fig:example_figure}
\end{figure}

\section{Force-free regime}
After estimating the energy of the particles accelerated in Crab pulsar, we shall consider the energy particles possess, at the edge of the Crab Nebula. We will consider a case, in which particles move long the Archimedes spirals, meaning that curved trajectory of the particles is
\begin{equation}
\phi (r) = -a r
\end{equation}
with $a$ being constant. It is shown, that in the limit of large distance from the rotator, the velocity of the particles becomes constant. Which, on the other hand, means, that particles are on force-free regime. 

It is truly very important, that particles maintain force-free regime. In our consideration, force-free regime is the key for particle energies to remain constant during the process of leaving the pulsar and entering the Nebula.

This assumption, gives us the reason to think, that if the particles move on the curves like Archimedes spirals, they can maintain force-free regime and leave the outer parts of the Crab Nebula without any loss in their energy. This way of thinking makes it quite possible, that the energy of the particles from the observed high energy - Tev emissions, are gained in the vicinity of the Crab Pulsar and emitted from the outer parts of the Crab Nebula, without any loss. 

\section{Possible radiation mechanisms}
\label{sec:maths} % used for referring to this section from elsewhere

Previously experienced flaring episodes coming from Crab Nebula, are believed to be explained by the Inverse-Compton mechanism (Aliu, 2014). Two weeks lasting flaring episode from the Crab Nebula first was detected in 2013. At this point, it was concluded, that synchrotron emission of relativistic electrons was mechanism explaining the phenomenon, but for higher energies, it was considered to be explained by the inverse-Compton upscattering mechanism. But the cause of the flares still remained unexplained by any mechanism. First, we want to examine, inverse-Compton mechanism for the data made by observations.
And we will start with considering an inverse-Compton upscattering process, with output frequencies 
\begin{equation}
\nu' = {\nu{{1+{v\over{c}}\cos{\theta}}\over{1-{v\over{c}}\cos{\theta'}+{{{h \nu}\over{\gamma m c ^ 2}} {(1-\cos{\theta}-\cos{\theta'}})}}}}
\end{equation}
where $\nu'$ is the output frequency, $\theta$ is the angle between incident photon and electron and $\theta'$ is the angle between reflected photon and the electron. 

Equation (7) constains Klein-Nishina term ${{{h \nu}\over{\gamma m c ^ 2}} {(1-\cos{\theta}-\cos{\theta'}})}$, which become important only when

\begin{equation}
{{h \nu}\over{\gamma m c ^ 2}} >> 1
\end{equation}

Which means, that mass of the electron is much less than the mass of the photon. If this condition is true, photon scattering won't take place in plasma. Thus, we can disregard Klein-Nishina term.

We consider, that output frequency corresponds to Tev energies. Assuming, that the process is explained by the inverse-Compton upscattering mechanism we tried to evaluate the range of possible angles
\begin{figure}
	\includegraphics[width=\columnwidth]{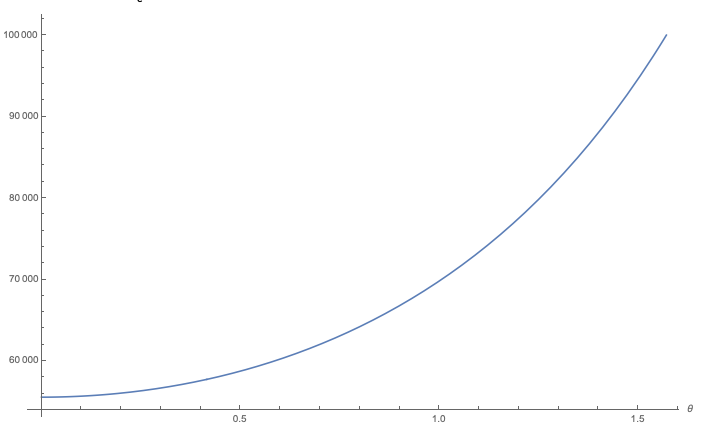}
    \caption{Taking into account the input and the output frequencies of the incident photon on the electron, we evaluated the angle on the reflected photon. The graph shows, that for $\nu=10^{2}$ and the observed output frequency $\nu'=10^{5}$ the range of the angle $\theta'$ between reflected photon and the electron seemed to be very little.}
    \label{fig:example_figure}
\end{figure}
Thus for $\nu=10^{2}$ and the observed output frequency $\nu'=10^{5}$ the range of the angle $\theta'$ between reflected photon and the electron seemed to be very little. Which led us to conclude, that probability of the mechanism explaining the phenomenon, being inverse-Compton upscattering is very little. To go further, we checked how the incident and reflecting angles $\theta'$ and $\theta$ were depended on each other.
\begin{figure}
	\includegraphics[width=\columnwidth]{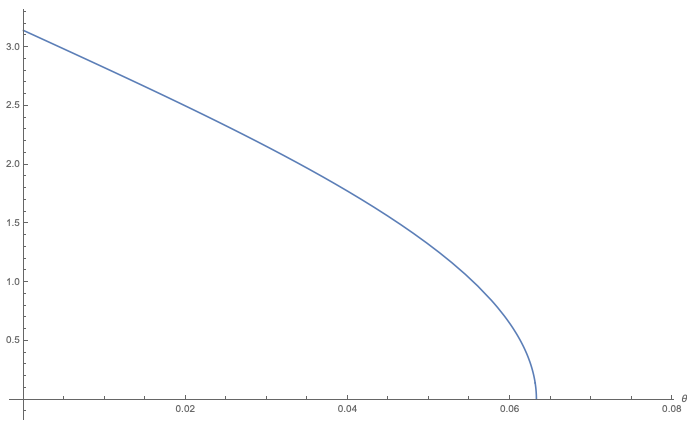}
    \caption{We checked how the incident and reflecting angles $\theta'$ and $\theta$ were depended on each other. the graph shows, that the incident and the reflecting photon angles have very little ranges. Which means, that the incident and the reflecting photon should almost be aligned. This also proves, that Inverse-Compton upscattering mechanism has very little probability}
    \label{fig:example_figure}
\end{figure}
This also proves, that Inverse-Compton upscattering mechanism has very little probability. To make it more clear, we think, that observed flaring episodes of Crab Nebula, can be explained with inverse-Compton upscattering mechanism. Only it has very little probability, which can easily be calculated from the assumptions and calculations we have made.

Due to very high energies of the particles Inverse-Compton mechanism is defines by Klein-Nishina regime. Klein-Nishina time-scale of the process $t_{KN} = E/P_{KN}$. Using the fact, that particle energy $E_{p} = 100 Tev$, derived time-scale will exceed the typical time-scale by many orders.

%We consider plasma cloud, containing big amount of Debye volumes. The system with particles and Langmuir waves is considered as quasi-stationary, as the relativistic particles leave the volume, but they are substituted by other identical ones.

%Now, if we assume, that electromagnetic wave energy is less than the energy of the Langmuir waves and incident electromagnetic wave causes spatial shifts within the multitude of dipoles, we are able to write down restoring force, for displaced charged particles. For general case, we take it to have nonlinear part as well

%\begin{equation}
%{\bf f}(t) = - \eta {\bf r}(t) - q {\bf r}^{3}(t)
%\end{equation}

%Later after writing the equation of motion

%\begin{equation}
%m \gamma^{3} \frac{d^{2} {\bf r}}{dt^{2}} = - m \gamma \Gamma \frac{d{\bf r}}{dt} - %\eta {\bf r} - q {\bf r}^{3} + e {\bf E}
%\end{equation}

%for polarization vector
%{\bf P}  we derive:
%\begin{equation}
% \frac{d^{2} {\bf P}}{dt^{2}} + \Gamma \frac{d {\bf P}}{dt} +
% \left(\frac{\omega_0}{\gamma} \right)^{2} {\bf P}
% + Q {\bf P}^{3} = {\left(\frac{e^{2} N_D}{m \gamma^{3}} \right)} {\bf E}
%\end{equation}
%where $Q \equiv q/m e^{2} N_{D}^{3} \gamma^{3}$ and $\omega_0$, in
%this case, is connected to the Langmuir oscillation frequency $\omega_{0} =
%\omega_{p}/\gamma^{3/2}$.

\section*{Discussion}
We have considered a few models of particle acceleration in Crab pulsar. Landau damping of Langmuir waves can explain particles having such high energies. The crucial part of the phenomenon involves force-free regime. We argue, that high energy particle are able to enter the Nebula without any loss in their energetic supply, that can be maintained by moving on special curves. In the later sections we discussed the possibility of Inverse-Compton up-scattering mechanism in Crab pulsar.

We have also considered radiation mechanisms, which accelerate particles up to Tev energies and these particles then maintain force-free regime.

\section*{Acknowledgments}
The work was supported by Shota Rustaveli National Science Foundation of Georgia (SRNSFG) [FR18\_564 ].

\bibliographystyle{plain} 
\bibliography{bibliography.bib} 

% Don't change these lines
\bsp    % typesetting comment
\label{lastpage}
\end{document}